\documentclass[runningheads]{svmult}

\usepackage{makeidx}   
\usepackage{graphicx}  
\usepackage{subeqnar}  
\usepackage{multicol}  
\usepackage{physprbb}  
\makeindex             

\newcommand{\lesssim}{\mathbin{\lower 3pt\hbox
   {$\rlap{\raise 5pt\hbox{$\char'074$}}\mathchar"7218$}}} 
\newcommand{\gtrsim}{\mathbin{\lower 3pt\hbox
   {$\rlap{\raise 5pt\hbox{$\char'076$}}\mathchar"7218$}}} 

\begin{document}
\title*{Fundamental Physics with the SKA: \\
Strong-Field Tests of Gravity Using Pulsars and
Black Holes}
\toctitle{Fundamental Physics with the SKA: 
\protect\newline
Strong-Field Tests of Gravity Using Pulsars and
Black Holes}
%
%
\titlerunning{Strong-Field Tests of Gravity
}
%
\author{Michael Kramer}
\authorrunning{Michael Kramer}
%
%
\institute{University of Manchester, Jodrell Bank Observatory,
Cheshire SK11 9DL, UK}

\maketitle              

\begin{abstract}
The Square-Kilometre-Array (SKA) will be a radio telescope with a
collecting area that will exceed that of existing telescopes by a
factor of a hundred or so. This contribution summarises one of the
key-science projects selected for the SKA.

The sensitivity of the SKA allows us to perform a Galactic Census of
pulsars which will discover a large fraction of active pulsars beamed
to us, including the
long-sought for pulsar-black hole systems. These systems are unique in
their capability to probe the ultra-strong field limit of relativistic
gravity. By using pulsar timing we can determine the properties of
stellar and massive black holes, thereby testing the
Cosmic Censorship Conjecture and the No-Hair theorem. The large
number of millisecond pulsars discovered with the SKA will also
provide us with a dense array of precision clocks on the sky.  These
clocks will act as the multiple arms of a huge gravitational wave
detector, which can be used to detect and measure the stochastic
cosmological gravitational wave background that is expected from a
number of sources.
\end{abstract}

\section{Introduction}
\label{einstein}

Solar system tests provide a number of very stringent tests of
Einstein's theory of general relativity (GR), and to date GR has
passed all observational tests with flying colours.  Nevertheless, the
fundamental question remains as to whether Einstein has the last word
in our understanding of gravity or not. Solar-system experiments are
all made in the weak-field regime and will never be able to provide
tests in the strong-field limit which is largely unexplored.  Tests
involving the observations of X-ray binaries may help, but the
interpretation of these results depends to some extent on models which
are known with only limited precision. In contrast, pulsars represent
accurate clocks which can, in a binary orbit, allow us to perform high
precision tests of gravitational theories. In the following we
describe a key-science project developed for the SKA with a number of
colleagues, namely Don Backer, Jim Cordes, Simon Johnston,
Joe Lazio and Ben Stappers.  Details can be found in
\cite{kbc+04,ckl+04}.


\section{Strong-field tests of gravity}

Through its sensitivity, sky and frequency coverage, the SKA will
discover a very large fraction of the pulsars in the Galaxy, resulting
in about 20,000 pulsars.  This number represents essentially all
active pulsars that are beamed toward Earth and includes the
discovery of more than 1,000 millisecond pulsars (MSPs).  This
impressive yield effectively samples every possible outcome of the
evolution of massive binary stars.  The sensitivity of the SKA allows
much shorter integration times, so that searches for compact binary
pulsars will no longer be limited. Among the discovered sources,
pulsar-black hole (PSR-BH) systems are to be expected. Being timed
with the SKA, a PSR-BH system would be an amazing probe of
relativistic gravity with a discriminating power that surpasses all of
its present and foreseeable competitors \cite{de98}.

\subsection{Black Hole properties}

As stars rotate, astrophysicists also expect BHs to rotate, giving
rise to both a BH spin and quadrupole moment. The resulting
gravito-magnetic field causes a relativistic
frame-dragging in the BH vicinity, leading the orbit of any test mass
about the BH to precess if the orbit deviates from the equatorial
plane. The consequences for timing a pulsar around a BH
have been studied in detail by Wex \& Kopeikin (1999 \cite{wk99}), who
showed that the study of the orbital dynamics allows us to use the
orbiting pulsar to probe the properties of the rotating BH.
Not only can the mass of the BH be measured with very high accuracy,
but the spin of the BH can also be determined precisely using the
nonlinear-in-time, secular changes in the observable quantities due to
{\em relativistic} spin-orbit coupling. The anisotropic nature of the
quadrupole moment of the external gravitational field will produce
characteristic short-term periodicities due to {\em classical}
spin-orbit coupling, every time the pulsar gets close to the oblate BH
companion \cite{wex98,wk99}. Therefore, the mass, $M$, and both the
dimensionless spin $\chi$ and quadrupole $q$,
\begin{equation}
   \chi \equiv \frac{c}{G}\; \frac{S}{M^2} \qquad \mbox{\rm and} \qquad
   q = \frac{c^4}{G^2}\; \frac{Q}{M^3} 
\end{equation}
of the BH can be determined, where $S$ is the angular momentum and $Q$
the quadrupole moment. These measured properties of a BH can be
confronted with predictions of GR.

In GR, the curvature of space-time diverges at the
centre of a BH, producing a singularity, which physical behaviour is
unknown. The Cosmic Censorship Conjecture was invoked by Penrose in
1969 (see e.g.~\cite{hp70}) to resolve the fundamental concern that if
singularities could be seen from the rest of space-time, the resulting
physics may be unpredictable. The Cosmic Censorship Conjecture
proposes that singularities are always hidden within the event
horizons of BHs, so that they cannot be seen by a distant
observer.  A singularity that is found not to be hidden but ``naked''
would contradict this Cosmic Censorship. In other words, the complete
gravitational collapse of a body always results in a BH rather
than a naked singularity (e.g.~\cite{wal84}).
We can test this conjecture by measuring the spin of a
rotating BH: In GR we expect $\chi\le1$.  If, however, SKA
observations uncover a massive, compact object with $\chi>1$, two
important conclusions may be drawn. Either we finally probe a region
where GR is wrong, or we have discovered a collapsed object where the
event horizon has vanished and where the singularity is exposed to the
outside world. 

One may expect a complicated relationship between the spin of the BH,
$\chi$, and its quadrupole moment, $q$. However, for a rotating Kerr BH
in GR, both properties share a simple, fundamental relationship
\cite{tpm86}, i.e.~$q = -\chi^2$.
This equation reflects the ``No-hair'' theorem which implies
that the external gravitational field of an astrophysical (uncharged)
BH is fully determined by its mass and spin.
Therefore, by determining $q$ and $\chi$ from
timing measurements with the SKA, we can confront this fundamental
prediction of GR for the very first time.

The best timing precision would be provided by a PSR-BH system with a
MSP companion. Such systems do not evolve in standard scenarios, but
they can be created in regions of high stellar density due to exchange
interactions. Prime survey targets would therefore be the innermost
regions of our Galaxy and Globular Clusters. Finding pulsars in
orbits around massive or super-massive BHs would allow us to apply the
same techniques for determining their properties as for the stellar
counterpart \cite{wk99}.  Since the spin and quadrupole moment of a BH
scale with its mass squared and mass cubed, respectively, relativistic
effects would be measured much easier.

\begin{figure}[b]
\begin{center}
\includegraphics[width=.5\textwidth]{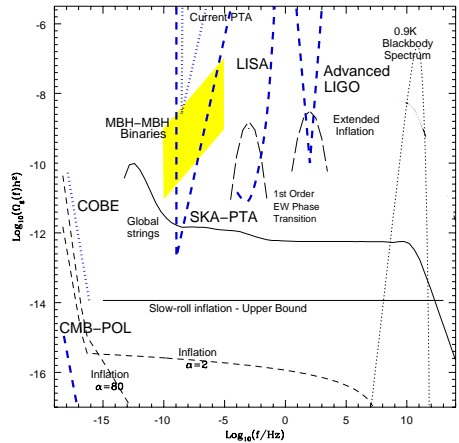}
\end{center}

\caption{\label{fig:gwspec}
Summary of the potential cosmological sources of a stochastic
gravitational background
as presented  Battye \& Shellard (1996) and sensitivity
curves for various experiments (see text).}
\end{figure}
\nocite{bs96}


\subsection{Gravitational Wave Background}
\label{sec:level0gw}

The SKA will discover a dense array of MSPs distributed across
the sky. Being timed to very high precision ($<$100 ns), this ``Pulsar
Timing Array'' (PTA) acts as a cosmic gravitational wave (GW)
detector.  Each pulsar and the Earth can be considered as free masses
whose positions respond to changes in the space-time metric.  A
passing gravitational wave perturbs the metric and hence affects the
pulse travel time and the measured arrival time at Earth
\cite{det79,fb90,rr95a}.  With observing times of a few years, pulsars
are sensitive to GWs frequencies of $f>1/T$, hence in the $\sim$nHz
range. Consequently, the SKA can detect the signal of a stochastic
background of GW emission in a frequency range that is 
complementary to that covered by LISA and LIGO.

A stochastic gravitational wave background should arise from a variety of
sources. Cosmological sources include inflation, string cosmology,
cosmic strings and phase transitions (see
Figure 1).
We can write the intensity of this GW background as
\begin{equation}
\Omega_{\rm gw}(f)=\frac{1}{\rho_c}\frac{d\rho_{gw}}{d\log{f}}
\end{equation}
where $\rho_{\rm gw}$ is the energy density of the stochastic
background and $\rho_c$ is the present value of the critical energy
density for closure of the Universe, $\rho_c={3H_0^2}/{8\pi G}$ with
$H_0\equiv h_0\times 100$ km s$^{-1}$ Mpc$^{-1}$ as the Hubble
constant. A contribution to the GW background is also expected from
astrophysical processes, i.e.~the coalescence of massive BH binaries
during early galaxy evolution \cite{rr95a,jb03}.  The amplitude of
this signal depends on the mass function of the massive BHs and their
merger rate \cite{jb03}. Measuring the slope of the spectrum would
allow us to discriminate between this foreground signal and the
cosmological sources.  The wedge-like sensitivity curve of the PTA is
shown in Fig.~\ref{fig:gwspec}. For timing precision that is only
limited by radiometer noise, the RMS is expected to scale with the
collecting area of the observing telescope.  In reality, the precision
is also affected by other effects.  Their limiting influence and the
application of correction schemes will need to be determined on a case
by case basis. However, extrapolating from the experience with the
best performing MSPs today, we can expect the SKA to improve on the
current limit on $h_0^2\Omega_{\rm gw}$ by a factor $\sim10^4$!


\end{document}